# Scaling of laser-driven ion energies in the relativistic transparent regime

D. Jung[1,2], L. Yin[1], B. J. Albright[1], D. C. Gautier[1], B. Dromey[2], R. Shah[1], S. Palaniyappan[1], S. Letzring[1], H.-C. Wu[1], T. Shimada[1], R. P. Johnson[1], D. Habs[4], M.Roth[3], J. C. Fernandez[1], B. M. Hegelich[1]

[1] Los Alamos National Laboratory, Los Alamos, New Mexico 87545, USA
[2] Queen's University Belfast, Belfast BT7 1NN, UK
[3] Technische Universität Darmstadt, 64289 Darmstadt, Germany
[4] Department für Physik, Ludwig-Maximilians- Universität München, D-85748 Garching, Germany

**Laser-driven ions have compelling properties and their potential use for medical applications has attracted a huge global interest. One of the major challenges of these applications is generating beams of the required energies. To date, there has been no systematic study of the effect of laser intensity on the generation of laser-driven ions from ultrathin foils during relativistic transparency. Here we present a scaling for ion energies with respect to the on-target laser intensity and in considering target thickness we find an optimum thickness closely related to the experimentally observed relativistic transparency. A steep linear scaling with the normalized laser amplitude $a_0$ has been measured and verified with PIC simulations. In contrast to TNSA, this scaling is much steeper and has been measured for ions with Z > 1. Following our results, ion energies exceeding 100MeV/amu are already accessible with currently available laser systems enabling realization of numerous advanced applications.**

For more than a decade, intense short pulse lasers have been used to drive energetic ion beams [1–7]. These laser-driven ion beams have high particle numbers, energies up to several tens of MeV/amu [8] and a low transverse emittance [9] and a promise to be a competitive alternative to conventional radio frequency accelerators. The range of applications covers medicine with hadron cancer therapy [10], threat reduction (e.g. detection of fissile material) [11, 12] or energy generation with concepts in ion fast ignition [13–15]. The major drawback so far is that ion energies are typically too low for any of these applications. For the hadron cancer therapy, protons of 250MeV or carbon $C^{6+}$ ions of 4–5 GeV are needed. In the regime of the Target Normal Sheath Acceleration (TNSA) [1–3] protons have been accelerated to 67MeV [16] with laser intensities exceeding $10^{20}$W/cm$^2$. Heavier ions have merely reached several MeV/amu due to protons shielding the accelerating fields in this mechanism. Even with target cleaning, energies have not passed 10 MeV\amu [6]. Recent advances in laser intensities and contrast enabled exploration of new acceleration mechanisms such as the radiation pressure acceleration (RPA) [17–21] or the Break-Out Afterburner (BOA) [22–25] mechanism, which reach higher ion energies for both protons and heavier ions.

Scaling laws for ion energies are a fundamental requirement for the design of future laser systems and the realization of advanced applications. So far, scaling laws have been discussed for the TNSA mechanism [5, 7] and also for RPA dominated acceleration [18, 27, 41]. Here, we extend this research framework by presenting experimental data and theoretical analysis for scaling of ion energies in the relativistic transparent regime, where the BOA mechanism is operative. In a series of experiments we investigated how maximum ion energies scale for the BOA. Ion energies have been measured for intensities of $8 \times 10^{19}$, $2 \times 10^{20}$ and $1.7 \times 10^{21}$W/cm$^2$ at the Los Alamos National Laboratory using the Trident laser system [28] (see *Methods*). Expressed in units of the normalized laser amplitude $a_0$, these intensities equal $a_0$=8.6, 14.7 and

34.5, respectively ($a_0 = I_0[W/cm^2] * \lambda^2[\mu m]/1.3718$). In laser-matter interaction with a normalized laser amplitude $a_0 > 1$, electrons gain relativistic kinetic energies in the laser field and the interaction is thus commonly referred to as being *relativistic*. At these intensities -if parameters are chosen accordingly- the target undergoes a phase of relativistic transparency during the peak laser pulse interaction.

Relativistic transparency defines a special state of the plasma, where it is classically overdense, yet relativistically underdense. Classically, i.e. neglecting relativistic effects, the normalized electron density $N = n_e/n_{cr} > 1$ with the critical electron density $n_{cr} = m_e \omega_\lambda^2/(4\pi e^2)$ with $\omega_\lambda$ the laser frequency. Accounting for the electron mass increase by the Lorentz factor $\gamma_e$ due to their relativistic energies, however, $N' \approx N/\gamma_e \leq 1$ and the plasma is relativistically transparent. Using the Trident laser we found that with intensities exceeding $5 \times 10^{19} W/cm^2$ diamond targets with thickness below 600 nm turn relativistically transparent during the peak laser interaction as recently reported by Palaniyappan, et al. [37].

When the plasma undergoes such a phase of relativistic transparency during the peak pulse interaction, efficient acceleration via the Break-Out Afterburner (BOA) is possible [23]. When the target is relativistically transparent, electrons are accelerated toward the rear of the target in the intense laser field, which sets up a strong longitudinal electric field that co-moves with the ions. During this phase the laser continuously imparts forward momentum to the electrons, which couple to the ions, thus accelerating them to extreme energies exceeding 100MeV/amu [38]. When the target eventually becomes classically underdense with $N \leq 1$, acceleration of ions decreases significantly due to the low coupling efficiency of laser light into the low-density plasma. In contrast, the target stays classically and relativistically overdense during acceleration in the TNSA and RPA regime. When transparency occurs, the plasma reflectivity $R(\omega)$ drops substantially, decreasing the efficiency of energy transfer through the radiation pressure significantly [41].

Ion spectra have been measured using an ion wide angle spectrometer (as described in Ref.[29], see *Methods*)) and thin, free-standing, artificially grown diamond foils [35] with density of ~ 3 g/cc have been used as target material. Thicknesses range from 30 nm to 5 µm; the bulk proton concentration is naturally very low and hydrogen-contamination on the surface is the main source of protons (see *Methods*). In order to find the maximum possible energy at a given intensity, the optimum thickness needs to be determined. This is necessary as dynamics of the plasma density evolution and thus the relativistic transparent phase strongly depend on the target and laser parameters. In particular, the laser contrast has a difficult to predict influence on these dynamics; premature heating of the target under the laser pedestal and the rising edge of the laser pulse lower the initial target density "seen" by the peak of the laser pulse.

Therefore, a full thickness scan has been conducted for each intensity to find the optimum thickness and maximum possible energy. Three thickness scans with on-target intensities ranging from $10^{19}$ to $10^{21} W/cm^2$ have been performed. In Fig. 1 a) typical maximum carbon $C^{6+}$ energies are shown as a function of target thickness (red stars: ~ $8 \times 10^{19} W/cm^2$ ($a_0 = 8$), blue circles: ~ $2 \times 10^{20} W/cm^2$ ($a_0 = 13$) and the green triangles: ~ $1 \times 10^{21} W/cm^2$ ($a_0 = 29$). The increasing scatter with increasing intensity is a result of the stronger fluctuations in the on-target intensities and stronger sensitivity to target alignment. This is mainly caused by the reduced Rayleigh length $z_r$ of the laser focus, which is proportional to the aperture of the different focusing optics used to control the intensity and additionally by shot-to-shot fluctuations in the laser contrast, especially at higher intensities. The optimum thickness for each intensity has been estimated using standard peak fitting (solid lines in Fig. 1 a)); it ranges from 75 nm to 175 nm for $8 \times 10^{19}$ W/cm$^2$ with peak at 130nm, from 100 nm to 300 nm for $2 \times 10^{20}$ W/cm$^2$ with peak at 230 nm and from 350 nm to past 550 nm for $\sim 1 \times 10^{21}$ W/cm$^2$ with peak at 480nm (where the latter fit has a higher uncertainty due to less data points and higher fluctuations). The peaks are shown in Fig. 1 b) as a function of the normalized laser amplitude (black stars) overlaid with the experimental data points within the found thickness ranges (red squares).

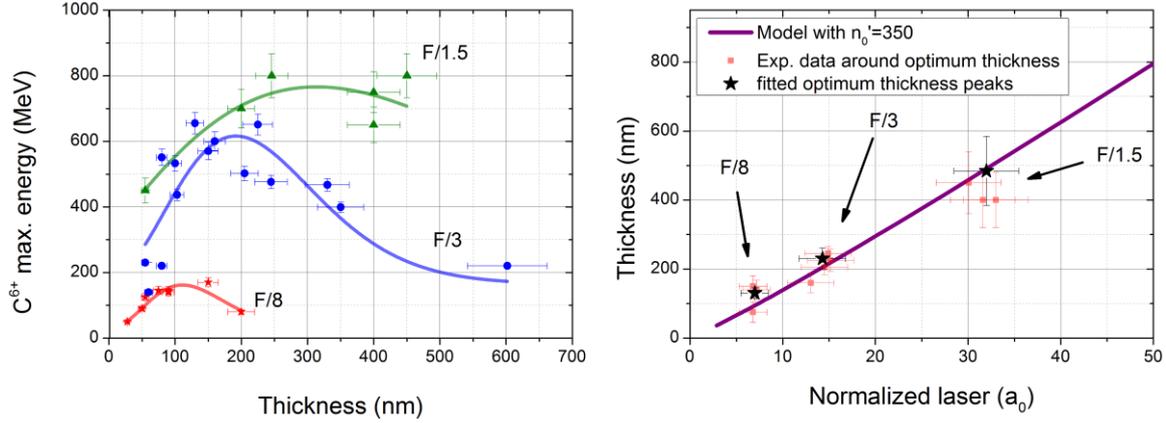

Figure 1 Left frame: Maximum carbon $C^{6+}$ energies. Red stars are for intensities of $\sim 8 \times 10^{19}$ W/cm$^2$ ($a_0$ = 8) obtained with an F/8 OAP, the blue circles for $\sim 2 \times 10^{20}$ W/cm$^2$ ($a_0$ = 13) obtained with the F/3 OPA (see also Ref.[42]) and the green triangles for $\sim 1 \times 10^{21}$ W/cm$^2$ ($a_0$ = 29) obtained with the F/1.5 OAP. Solid lines are fits using Giddings peak function to obtain the optimum thickness. Note that the increasing scatter with increasing intensity in the measured energies is a result of the stronger fluctuations in the on-target intensity and stronger sensitivity to target alignment due to the reduced Rayleigh length of the laser focus. Right frame: Plot showing data points around the optimum thickness (light red square) as a function of the normalized laser amplitude $a_0$. The black stars are the values obtained from the Giddings peak function fits from the left frame and the purple line is a plot of the optimum thickness calculated using the model presented in Eq(2) using an adjusted initial intensity $n_0'$ of 350.

In Fig. 2 a), only the highest measured energies for each intensity are plotted (red stars) as a function of the normalized laser amplitude $a_0$. We also added results published by Henig, et al. [24] (blue star) for ion acceleration in the relativistic transparent regime, also performed on Trident, but using plasma mirrors. For the intensities covered, the results have been fitted using a linear correlation between $a_0$ and the maximum ion energy (red solid line) with $E_{max}=28(a_0-0.7)$. (It should be noted that since we could only cover 3 intensities experimentally, power-laws ranging from 0.2 to 1 show similar confidences.) We also performed several 2D with parameters similar to the experiment and one large-scale 3D simulation to confirm that the salient BOA dynamics are contained in 2D. The simulations employ the two and three-dimensional, relativistic, electromagnetic, charge-conserving, PIC code VPIC [43] (see *Methods*). The results are shown in Fig. 2 b) where 2D simulations are denoted by "2" and the 3D simulation by "3". A linear regression of the PIC results for the covered range from 1 to 50 gives $E_{max}=33(a_0-0.8)$ (blue solid line). The results are in good agreement with our experimental data as shown in Fig. 2 c).

In order to develop an analytic expression for the scaling law and the optimum thickness to be used as a tool with predictive capabilities, we invoke the model published by Yan, et al. [44], which is based on the electron reflexing model by Mako and Tajima [45]. In this model, acceleration occurs between the onset of the relativistic transparency at time $t_1$ when the target has turned relativistically transparent (when the density has decreased to $N/\gamma \leq 1$) and the time $t_2$ when the target has turned classically underdense ($N \leq 1$ and $N/\gamma \ll 1$, see *Methods* for details). In this model the maximum energy is calculated through the response of the

(nonrelativistic) ions to the electrostatic field, i.e., to the time averaged electron energy $E_0 = m_e c^2/\Delta t \int((a^2(t') + 1)^{1/2}-1)dt'$ due to the ponderomotive force and the duration $\Delta t = t_2 - t_1$ of the relativistic transparent phase. In Fig. 2 b), $E_{max}$ has been plotted for Trident parameters with different laser focus sizes (green circles, 80 J, $\tau_\lambda$ = 600 fs FWHM, a(t') with a $\sin^2$ temporal envelope). The model predictions match the 2D and 3D-VPIC and experimental results.

Taking the ansatz that $E_{max} \propto a_0 \tau_\lambda^x$, a parametric study over a large parameter range gives x = 0.28 and the following expression for the scaling

$$E_{max} \approx 5\tau_\lambda^{0.28} (a_0 - 1) MeV \propto I_L^{1/2} \tau_\lambda^{0.28} \quad , \tag{1}$$

where $\tau_\lambda$ is in units of femtoseconds. In Fig.2 b) the scaling has been plotted using $\tau_\lambda$ = 600 fs. This yields for Trident-like parameters a maximum energy $E_{max}$ = 30MeV $\ast (a_0 - 1)$ according to the analytical model. It is apparent that the expression only applies for relativistic laser intensities with $a_0 \geq 1$. The curve is in very good agreement with the simulations and the experimental data as shown in Fig. 2 c). Using $\tau_\lambda$ = 45 fs and $a_0$ = 5, the model yields $E_{max}$ = 60MeV, which is close to the maximum carbon energies observed in an experiment published by Steinke, et al [46] (with a target thickness of 7 nm basic conditions for relativistic transparency during the interaction are given for these parameters).

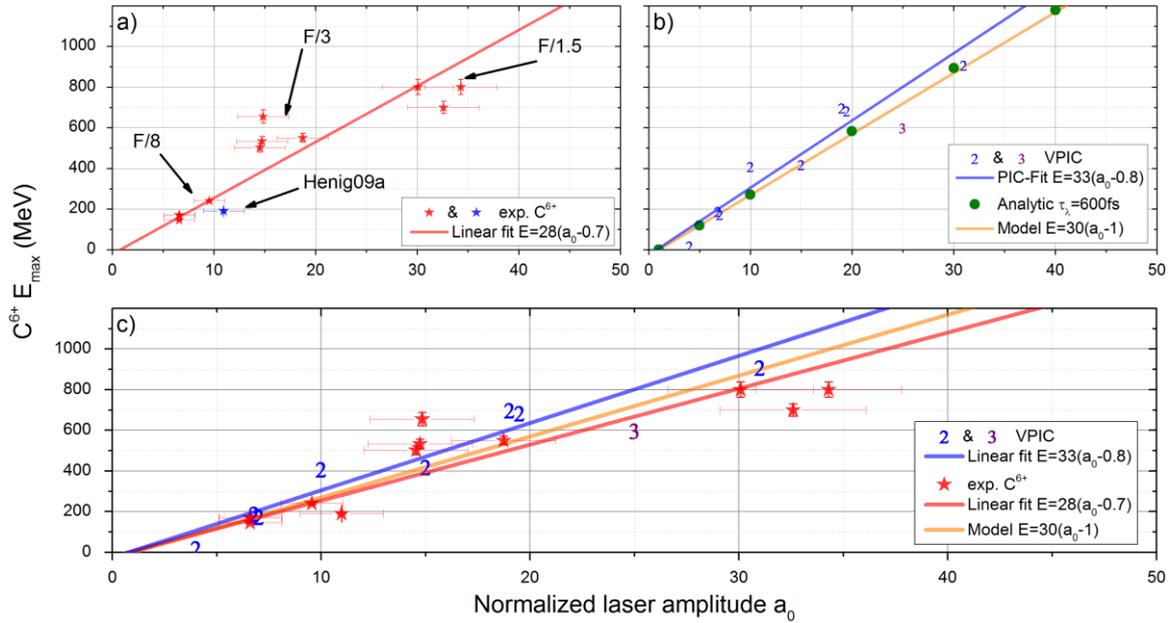

Figure 2 a) Experimental results for maximum carbon $C^{6+}$ ion energies as a function of the normalized laser amplitude $a_0$ (red stars). Intensities of $8\times10^{19}$, $2\times10^{20}$ and $1\times10^{21}$ W/cm² have been generated using different final focusing optics with F-number F/8, F/3, F/1.5, respectively. The experimental data indicates a linear dependency between the maximum laser energy and the normalised laser amplitude for the covered intensity range. b) Analytical (green circles and yellow solid line) )and PIC simulation results for 2D (blue 2) and 3D (purple 3) simulations with parameters close to the experiment. The analytic expression shows a linear dependency between the maximum energy and the normalized amplitude and is in good agreement with the PIC simulations. c) Combined plot showing agreement between experimental and simulation result.

It should be noted here that this basic model only assumes a single species and charge and does not include any physics related to multi-species targets or ionization. In model by Yan, et al., the maximum energy is linearly dependent on the ion charge $q_i$ (see Methods) so that the leading factor 5 in Eq. 1 might be replaced with $5q_i/6$. In that sense, the model is strictly applicable only to the species experiencing the BOA dynamics during the relativistic transparent phase of the interaction and acceleration in the co-moving electric field. For the targets used here, this does not apply to the (surface-)protons, as they are removed from the diamond targets long before the peak pulse interaction [12, 23] due to "self-cleaning"[22]. Data collected for proton acceleration from CH-plastic targets with a large bulk-proton concentration presented by Hegelich and Jung, et al.[48] support the scaling presented here for protons.

Applying the scaling law to a given set of laser parameters also requires that the optimum target thickness is used. We have used the same technique as above to derive an analytic expression for the optimum thickness as

$$d_{opt} \approx 9.8 \times 10^{-10} I_L^{13/24}/N_0' \cdot \tau_\lambda[fs] \approx 6.7\ \tau_\lambda a_0^{13/12}/N_0' \approx 7\ a_0 \tau_\lambda/N_0'. \quad (2)$$

A drawback of this expression is its dependency on the target density $N_0'$ that is present at the arrival of the peak pulse. Laser contrast is not included in the model and $N_0'$ is thus a free parameter. Setting $N_0' = 400$ (half of the initial density) reproduces the experimentally observed optimum thicknesses for all three intensities. Using half the initial density for Steinke, et al [46] with $N_0' = 250$ in fact reproduces 7 nm as optimum thickness for their parameters.

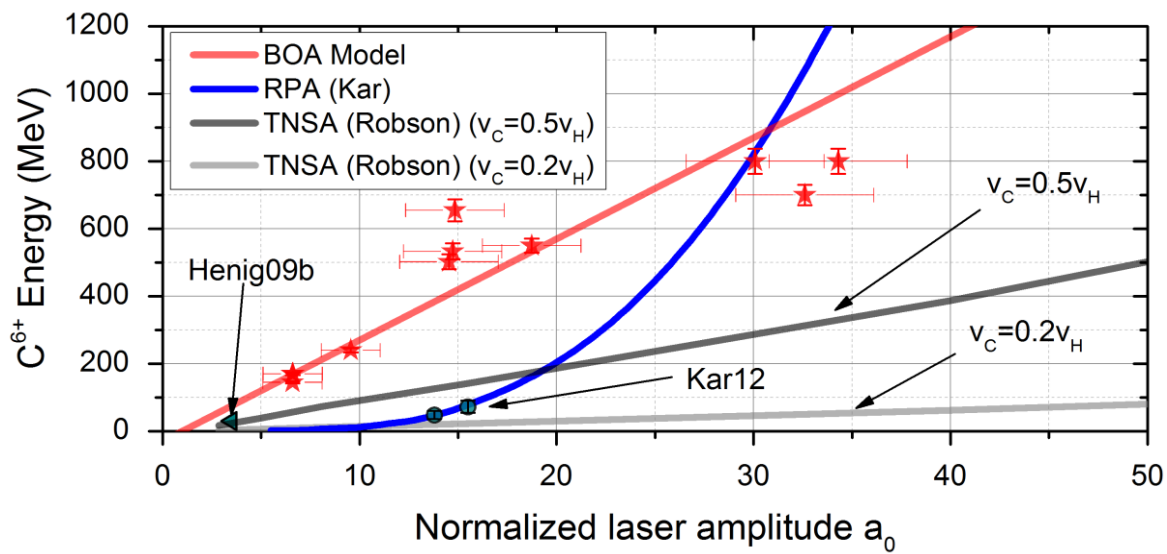

*Figure 3 Comparison of scaling models. BOA (solid green for $\tau_\lambda$ = 600 fs) and RPA for thickness of d = 100 nm (solid blue) with experimental data (red stars). The scaling for TNSA is adapted to carbon ions assuming velocity ratios to protons of 0.2 (light solid grey) and 0.5 (dark solid grey). Here, for the sake of simplicity, only maximum energies are compared; efficiency or spectral shape is not considered. Added are also experimental results for maximum energies measured in the RPA regime by Henig, et al. [50] and Kar, et al. [27].*

In Fig. 3, the scaling derived for acceleration via the BOA (red solid line) is compared with the two phase model with 3D effects for TNSA (grey solid lines) published by Robson, et al. [7] and the RPA light sail scaling (blue solid lines) in Kar, et al. [27] for similar laser parameters. It should be emphasized that here, for the sake of simplicity, only maximum energies are compared; efficiency into these energies or the spectral shape is not considered (which can be hugely different for the different acceleration mechanisms). For the TNSA scaling, we assume that acceleration is equally efficient for carbon ions as for protons and that carbon velocities attained in TNSA are a factor 0.5 slower than the respective proton energies due to their lower charge to mass ratio of 0.5 (dark grey solid line). However, so far the highest experimentally measured velocity ratio in TNSA for carbon ions to protons has only been 0.2 (light grey solid line) despite prior target cleaning [8], i.e. removal of surface protons. Even with optimistic conditions, the scaling with the laser intensity has a much smaller numerical coefficient (in TNSA regimes with planar targets where scaling is linear with $a_0$ [5]) than for the BOA. Comparison with RPA is less straight forward as the published scalings give energies for the centre energy of a quasi-monoenergetic spectrum; the scalings are also not optimized for thickness. To give a general idea of how BOA compares with RPA, we display the scaling using a set of laser and target parameters as in Kar, et al. [27] (blue solid line, Cu-target, 100 nm thickness, $\tau_\lambda$ = 800 fs). For non-relativistic ion energies, the scaling follows $a_0^4$; for relativistic ion energies, the dependency reduces to $a_0^2$ (where for fixed pulse durations $a_0$ is proportional to the fluence).

It should be noted that predictions of our scaling presented here are more uncertain for values of $a_0$ exceeding ~100 or intensities largely exceeding $10^{22}$W/cm$^2$. There is only very limited experimental data available for benchmarking predictive capabilities of our PIC simulations at these intensities. The presented comparison of the energy scalings clearly emphasizes the importance of these laws for design of future laser systems. For the parameters used here, for intensities below $a_0$ = 30 acceleration via BOA gives the highest energies. For higher intensities, RPA has the potential to exceed these energies, provided a 1D-geometry can be maintained [21, 49] and the target does not undergo excessive heating turning it relativistically transparent prior to arrival of the main pulse.

Acknowledgments
We are grateful for the support of the Trident laser team. The VPIC simulations were run on the LANL Roadrunner supercomputer. Work was supported by: DOE OFES, Los Alamos Laboratory Directed Research and Development program, Deutsche Forschungsgemeinschaft (DFG) Transregio SFB TR18, Cluster of Excellence (MAP). Work performed under the auspices of the U.S. Dept. of Energy by the Los Alamos National Security, LLC, Los Alamos National Laboratory.

**Methods**

*Laser:*
Trident has a typical pulse duration of 600 fs with 80 J of energy at a wavelength of 1.053 μm. Trident's high temporal laser contrast of $10^{-7}$ at -4 ps enables interaction of the peak pulse with an highly overdense target even for nm-thicknesses. The on-target intensity has been controlled using Off-axis parabolic (OAP) mirrors with F-numbers of F/8, F/3 and F/1.5 (as F/N with N=f/D where f is the focal length and D the diameter of the laser beam). The spot diameter is then proportional to Nλ with λ the laser wavelength. The diameters have been measured to be 13.8 μm (F/8), 7.9 μm (F/3) and 3.4 μm (F/1.5) with an encircled energy of 50%.

*Diagnostics:*
The iWASP is based on particle deflection in a magnetic field of order 0.5T and a 30 μm slit aperture to measure the spectra angularly resolved in one plane, covering a range of about 25°. Species separation is achieved by use of the hugely different stopping power of carbon ions and protons in a stacked detector consisting of a 32 µm Al layer (for laser light protection), followed by a 1mm thick CR39 [30] and a BAS-TR image plate (IP) [31, 32]. The CR39 detects carbon ions above 33MeV (the energy needed to pass the Al layer); protons above 11MeV leave no visible tracks on the CR39 (unless etched for several hours); here, typical etching times of 10 minutes have been applied. The IP is used to detect protons above 11MeV passing the CR39 in front of the IP. The energy resolution (lower instrument limit) at 80MeV/amu for protons and carbon C6+ ions is approximately ±3% and ±6%, respectively. The advantage of using the iWASP is its large solid angle of > 0.1msr, which is 3-5 orders of magnitude more than typically covered by conventional Thomson parabolas and results in a much higher precision of measuring the absolute peak ion energy. This is important especially for ion acceleration mechanisms, where maximum energies are emitted off-axis rather than on-axis. For the BOA mechanism it has recently been demonstrated in simulations and experimentally [33, 34] that maximum energies are in fact emitted off-axis, demanding a large solid angle for accurate measurements.

*Simulations:*
The 2D simulations use a domain of 50 µm×25 µm or 50 µm×50 µm in the (x, z) plane (the target transverse width is 25 µm or 50 µm). The laser pulse is polarized along y, propagates along x, and has a time-varying intensity profile $I(t) = I_0 \sin^2(t\pi/\tau)$, where $\tau/2$ = 540fs or 700fs is the FWHM. The central laser wavelength is 1054 nm, as in the experiments. The laser electric field has a 2D-Gaussian spatial profile with best focus at the target surface, where $E_y \sim \exp(-z^2/w^2)$ and w=5 µm. Solid density $C^{6+}$ (diamond like) targets at $n_e/n_{cr}$ = 660 or 821 (2.2 or 2.8 g/cm$^3$; $n_{cr} = m_e\omega_0^2/4\pi e^2$ is the critical density in CGS units and $\omega_0$ is the laser frequency) were employed with 5% or 20% protons in number density. The density is initially a constant-density slab profile. Both 2D and 3D (more details can be found in Yin, et. al. [33]) simulations retain the Debye-length-scale physics throughout the duration of the simulation. The size of the simulation box was chosen to give optimum resolution in the BOA regime where the acceleration region is localized to within the center of the target and does not depend sensitively on accurately capturing the electron dynamics outside this region [38].

*Model:*
The time $t_1$ is calculated via an 1D expansion of the target as $t_1 = (12/\pi^2)^{1/4}(N\tau d/(a_0C_s))^{1/2}$, with d the target thickness and ion sound speed $C_s \approx q_i m_e c^2 a_0/m_i$ and charge state $q_i$. Time $t_2$ is derived from a 3D isospheric expansion starting at $t_1$ with $t_2 = t_1 + Nd(\gamma^{1/3} - 1)/(\gamma C_s \sin(\omega t_1))$ and ω the laser frequency. The maximum energy is calculated via $E_{max} = (2\alpha + 1) q_i E_0 [ [1 + \omega_p(t_2 - t_1) ]^{1/(2a+1)} - 1]$ with $\omega_p$ the plasma frequency. The scaling factor α is the electron coherence parameter; for a wide range of parameters α has been found to be 3 in numerical PIC studies, provided $0.1 \leq Nd/a_0\lambda \leq 10$ (see Yan, et al.[44] for a detailed description and derivation of the model).

**References**
[1] S. P. Hatchett, C. G. Brown, T. E. Cowan, E. A. Henry, J. S. Johnson, M. H. Key, J. A. Koch, A. B. Langdon, B. F. Lasinski, R. W. Lee, A. J. Mackinnon, D. M. Pennington, M. D. Perry, T. W. Phillips, M. Roth, T. C. Sangster, M. S. Singh, R. A. Snavely, M. A. Stoyer, S. C. Wilks, and K. Yasuike, Phys. Plasmas 7, 2076 (2000).

[2] R. A. Snavely, M. H. Key, S. P. Hatchett, T. E. Cowan, M. Roth, T. W. Phillips,

M. A. Stoyer, E. A. Henry, T. C. Sangster, M. S. Singh, S. C.Wilks, A. MacKinnon,
A. Offenberger, D. M. Pennington, K. Yasuike, A. B. Langdon, B. F. Lasinski,
J. Johnson, M. D. Perry, and E. M. Campbell, Phys. Rev. Lett. 85, 2945 (2000).

[3] S. C. Wilks, A. B. Langdon, T. E. Cowan, M. Roth, M. Singh, S. Hatchett, M. H.
Key, D. Pennington, A. MacKinnon, and R. A. Snavely, Phys. Plasmas 8, 542
(2001).

[4] M. Borghesi, A. J. Mackinnon, D. H. Campbell, D. G. Hicks, S. Kar, P. K. Patel,
D. Price, L. Romagnani, A. Schiavi, and O. Willi, Phys. Rev. Lett. 92, 055003
(2004).

[5] J. Fuchs, P. Antici, E. d'Humieres, E. Lefebvre, M. Borghesi, E. Brambrink,
C. A. Cecchetti, M. Kaluza, V. Malka, M. Manclossi, S. Meyroneinc, P. Mora,
J. Schreiber, T. Toncian, H. Pepin, and P. Audebert, Nat Phys 2, 48 (2006).

[6] B. M. Hegelich, B. J. Albright, J. Cobble, K. Flippo, S. Letzring, M. Paffett,
H. Ruhl, J. Schreiber, R. K. Schulze, and J. C. Fernndez, Nature 439, 441 (2006).

[7] L. Robson, P. T. Simpson, R. J. Clarke, K.W. D. Ledingham, F. Lindau, O. Lundh,
T. McCanny, P. Mora, D. Neely, C.-G. Wahlstrom, M. Zepf, and P. McKenna, Nat.
Phys. 3, 58 (2007).

[8] M. Hegelich, S. Karsch, G. Pretzler, D. Habs, K. Witte, W. Guenther, M. Allen,
A. Blazevic, J. Fuchs, J. C. Gauthier, M. Geissel, P. Audebert, T. Cowan, and
M. Roth, Phys. Rev. Lett. 89, 085002 (2002).

[9] T. E. Cowan, J. Fuchs, H. Ruhl, A. Kemp, P. Audebert, M. Roth, R. Stephens,
I. Barton, A. Blazevic, E. Brambrink, J. Cobble, J. Fern´andez, J.-C. Gauthier,
M. Geissel, M. Hegelich, J. Kaae, S. Karsch, G. P. Le Sage, S. Letzring,
M. Manclossi, S. Meyroneinc, A. Newkirk, H. P´epin, and N. Renard-LeGalloudec,
Phys. Rev. Lett. 92, 204801 (2004).
[10] T. Tajima, D. Habs, and X. Yan, Rev. Accel. Sci. Tech. 2, 201 (2009).

[11] M. Roth, D. Jung, K. Falk, N. Guler, O. Deppert, M. Devlin, A. Favalli,
J. Fernandez, D. Gautier, M. Geissel, R. Haight, C. E. Hamilton, B. M. Hegelich,
R. P. Johnson, F. Merrill, G. Schaumann, K. Schoenberg, M. Schollmeier,
T. Shimada, T. Taddeucci, J. L. Tybo, F. Wagner, S. A. Wender, C. H. Wilde,
and G. A. Wurden, Phys. Rev. Lett. 110, 044802 (2013).

[12] D. Jung, K. Falk, N. Guler, O. Deppert, M. Devlin, A. Favalli, J. C. Fernandez,
D. C. Gautier, M. Geissel, R. Haight, C. E. Hamilton, B. M. Hegelich, R. P.
Johnson, F. Merrill, G. Schaumann, K. Schoenberg, M. Schollmeier, T. Shimada,
T. Taddeucci, J. L. Tybo, S. A. Wender, C. H. Wilde, G. A. Wurden, and M. Roth,
Phys. Plasmas 20, 056706 (2013).

[13] M. Tabak, J. Hammer, M. E. Glinsky, W. L. Kruer, S. C. Wilks, J. Woodworth,
E. M. Campbell, M. D. Perry, and R. J. Mason, Phys. Plasmas 1, 1626 (1994).

[14] M. Roth, T. E. Cowan, M. H. Key, S. P. Hatchett, C. Brown, W. Fountain,
J. Johnson, D. M. Pennington, R. A. Snavely, S. C. Wilks, K. Yasuike, H. Ruhl,
F. Pegoraro, S. V. Bulanov, E. M. Campbell, M. D. Perry, and H. Powell, Phys.
Rev. Lett. 86, 436 (2001).

[15] N. Naumova, T. Schlegel, V. T. Tikhonchuk, C. Labaune, I. V. Sokolov, and

G. Mourou, Phys. Rev. Lett. 102, 025002 (2009).

[16] S. A. Gaillard, T. Kluge, K. A. Flippo, M. Bussmann, B. Gall, T. Lockard, M. Geissel, D. T. Offermann, M. Schollmeier, Y. Sentoku, and T. E. Cowan, Phys. Plasmas 18, 056710 (2011).

[17] O. Klimo, J. Psikal, J. Limpouch, and V. T. Tikhonchuk, Phys. Rev. ST Accel. Beams 11, 031301 (2008).

[18] A. P. L. Robinson, P. Gibbon, M. Zepf, S. Kar, R. G. Evans, and C. Bellei, Plasma Physics and Controlled Fusion 51, 024004 (2009).

[19] X. Q. Yan, C. Lin, Z. M. Sheng, Z. Y. Guo, B. C. Liu, Y. R. Lu, J. X. Fang, and J. E. Chen, Phys. Rev. Lett. 100, 135003 (2008).

[20] A. Macchi, S. Veghini, and F. Pegoraro, Phys. Rev. Lett. 103, 085003 (2009).

[21] T. Esirkepov, M. Borghesi, S. V. Bulanov, G. Mourou, and T. Tajima, Phys. Rev. Lett. 92, 175003 (2004).

[22] L. Yin, B. J. Albright, B. M. Hegelich, K. J. Bowers, K. A. Flippo, T. J. T. Kwan, and J. C. Fernandez, Phys. Plasmas 14, 056706 (2007).

[23] B. J. Albright, L. Yin, K. J. Bowers, B. M. Hegelich, K. A. Flippo, T. J. T. Kwan, and J. C. Fernandez, Phys. Plasmas 14, 094502 (2007).

[24] A. Henig, D. Kiefer, K. Markey, D. C. Gautier, K. A. Flippo, S. Letzring, R. P. Johnson, T. Shimada, L. Yin, B. J. Albright, K. J. Bowers, J. C. Fern´andez, S. G. Rykovanov, H.-C.Wu, M. Zepf, D. Jung, V. K. Liechtenstein, J. Schreiber, D. Habs, and B. M. Hegelich, Phys. Rev. Lett. 103, 045002 (2009).

[25] B. Hegelich, D. Jung, B. Albright, J. Fernandez, D. Gautier, C. Huang, T. Kwan, S. Letzring, S. Palaniyappan, R. Shah, H.-C. Wu, L. Yin, A. Henig, R. Hrlein, D. Kiefer, J. Schreiber, X. Yan, T. Tajima, D. Habs, B. Dromey, and J. Honrubia, Nuclear Fusion 51, 083011 (2011).

[26] T. Esirkepov, M. Yamagiwa, and T. Tajima, Phys. Rev. Lett. 96, 105001 (2006).

[27] S. Kar, K. F. Kakolee, B. Qiao, A. Macchi, M. Cerchez, D. Doria, M. Geissler, P. McKenna, D. Neely, J. Osterholz, R. Prasad, K. Quinn, B. Ramakrishna, G. Sarri, O. Willi, X. Y. Yuan, M. Zepf, and M. Borghesi, Phys. Rev. Lett. 109, 185006 (2012).

[28] S. H. Batha, R. Aragonez, F. L. Archuleta, T. N. Archuleta, J. F. Benage, J. A. Cobble, J. S. Cowan, V. E. Fatherley, K. A. Flippo, D. C. Gautier, R. P. Gonzales, S. R. Greenfield, B. M. Hegelich, T. R. Hurry, R. P. Johnson, J. L. Kline, S. A. Letzring, E. N. Loomis, F. E. Lopez, S. N. Luo, D. S. Montgomery, J. A. Oertel, D. L. Paisley, S. M. Reid, P. G. Sanchez, A. Seifter, T. Shimada, and J. B. Workman, Rev. Sci. Instrum. 79, 10F305 (2008).

[29] D. Jung, R. Horlein, D. C. Gautier, S. Letzring, D. Kiefer, K. Allinger, B. J. Albright, R. Shah, S. Palaniyappan, L. Yin, J. C. Fernandez, D. Habs, and B. M. Hegelich, Rev. Sci. Instrum. 82, 043301 (2011).

[30] R. L. Fleischer, P. B. Price, and R. M. Walker, J. Appl. Phys. 36, 3645 (1965).

[31] I. J. Paterson, R. J. Clarke, N. C. Woolsey, and G. Gregori, Measurement Science and Technology 19, 095301 (2008).

[32] A. Mancic, J. Fuchs, P. Antici, S. A. Gaillard, and P. Audebert, Rev. Sci. Instrum. 79, 073301 (2008).

[33] L. Yin, B. J. Albright, K. J. Bowers, D. Jung, J. C. Fern´andez, and B. M. Hegelich, Phys. Rev. Lett. 107, 045003 (2011).

[34] D. Jung, L. Yin, D. C. Gautier, H.-C. Wu, S. Letzring, B. Dromey, R. Shah, S. Palaniyappan, T. Shimada, R. P. Johnson, J. Schreiber, D. Habs, J. C. Fernandez, B. M. Hegelich, and B. J. Albright, New J. Phys submitted (2013).

[35] "Applied diamond inc." [online] www.usapplieddiamond.com.

[36] V. A. Vshivkov, N. M. Naumova, F. Pegoraro, and S. V. Bulanov, Phys Plasmas 5, 2727 (1998).

[37] S. Palaniyappan, B. M. , Hegelich, H.-C. Wu, D. Jung, D. C. Gautier, L. Yin, B. J. Albright, R. P. Johnson, T. Shimada, S. Letzring, D. T. Offermann, J. Ren, C. Huang, R. Horlein, B. Dromey, J. C. Fernandez, and R. C. Shah, Nat. Phys. 8, 763 (2012).

[38] L. Yin, B. J. Albright, D. Jung, R. C. Shah, S. Palaniyappan, K. J. Bowers, A. Henig, J. C. Fernndez, and B. M. Hegelich, Phys. Plasmas 18, 063103 (2011).

[39] O. Buneman, Phys. Rev. 115, 503 (1959).

[40] K. J. Reitzel and G. J. Morales, Phys Plasmas 5, 3806 (1998).

[41] A. Macchi, S. Veghini, T. V. Liseykina, and F. Pegoraro, New J. Phys. 12, 045013 (2010).

[42] D. Jung, L. Yin, B. J. Albright, D. C. Gautier, S. Letzring, B. Dromey, M. Yeung, R. Horlein, R. Shah, S. Palaniyappan, K. Allinger, J. Schreiber, K. J. Bowers, H.-C. Wu, J. C. Fernandez, D. Habs, and B. M. Hegelich, New J. Phys. 15, 023007 (2013).

[43] K. J. Bowers, B. J. Albright, L. Yin, B. Bergen, and T. J. T. Kwan, Phys. Plasmas 15, 055703 (2008).

[44] X. Yan, T. Tajima, M. Hegelich, L. Yin, and D. Habs, Applied Physics B: Lasers and Optics 98, 711 (2010).

[45] F. Mako and T. Tajima, Physics of Fluids 27, 1815 (1984).

[46] S. Steinke, A. Henig, M. Schnrer, T. Sokollik, P. Nickles, D. Jung, D. Kiefer, R. Hrlein, J. Schreiber, T. Tajima, X. Yan, M. Hegelich, J. Meyer-ter Vehn, W. Sandner, and D. Habs, Laser and Particle Beams 28, 215 (2010).

[47] B. J. Albright, L. Yin, B. M. Hegelich, K. J. Bowers, T. J. T. Kwan, and J. C. Fern´andez, Phys. Rev. Lett. 97, 115002 (2006).

[48] B. M. Hegelich, D. Jung, B. J. Albright, M. Cheung, B. Dromey, D. C. Gautier,

C. Hamilton, S. Letzring, R. Munchhausen, S. Palaniyappan, R. Shah, H.-C. Wu, L. Yin, and J. C. Fernández, ArXiv e-prints (2013), arXiv:1310.8650 [physics.plasm-ph] .

[49] F. Dollar, C. Zulick, A. G. R. Thomas, V. Chvykov, J. Davis, G. Kalinchenko, T. Matsuoka, C. McGuffey, G. M. Petrov, L. Willingale, V. Yanovsky, A. Maksimchuk, and K. Krushelnick, Phys. Rev. Lett. 108, 175005 (2012).

[50] A. Henig, S. Steinke, M. Schnürer, T. Sokollik, R. Hörlein, D. Kiefer, D. Jung, J. Schreiber, B. M. Hegelich, X. Q. Yan, J. Meyer-ter-Vehn, T. Tajima, P. V. Nickles, W. Sandner, & D. Habs, Phys. Rev. Lett., 103, 245003 (2009)